\documentclass[modern]{aastex62}

\usepackage{graphicx}	% Including figure files
\usepackage{amsmath}	% Advanced maths commands
\usepackage{amssymb}	% Extra maths symbols
\usepackage{bm}
\usepackage{multirow}
\usepackage{comment}
\DeclareRobustCommand{\ion}[2]{%
\relax\ifmmode
\ifx\testbx\f@series
{\mathbf{#1\,\mathsc{#2}}}\else
{\mathrm{#1\,\mathsc{#2}}}\fi
\else\textup{#1\,{\mdseries\textsc{#2}}}%
\fi}

\def \Hb{H$\beta$~}
\def \OIII{[\ion{O}{iii}]~}
\def \deg{\hbox{$^\circ$}}
\def \kms{km~s$^{-1}$}
\def \ang{$\mathring{\mathrm{A}}$}
\submitjournal{ApJ}

\shorttitle{Outflows in SDSS~J1048+0055}
\shortauthors{Jaiswal et al.}

\begin{document}

\title{Jet-powered Outflows in Supermassive Black Hole Binary Candidate SDSS~J1048+0055}

%\maketitle
\correspondingauthor{Sumit Jaiswal}
\email{sumit@shao.ac.cn}

\correspondingauthor{Tao An}
\email{antao@shao.ac.cn}

\author{Sumit Jaiswal}
\affiliation{Shanghai Astronomical Observatory, Key Laboratory of Radio Astronomy, Chinese Academy of Sciences, 80 Nandan Road Shanghai 200030, China}

\author{Prashanth Mohan}
\affiliation{Shanghai Astronomical Observatory, Key Laboratory of Radio Astronomy, Chinese Academy of Sciences, 80 Nandan Road Shanghai 200030, China}

\author{Tao An}
\affiliation{Shanghai Astronomical Observatory, Key Laboratory of Radio Astronomy, Chinese Academy of Sciences, 80 Nandan Road Shanghai 200030, China}

\author{S\'{a}ndor~Frey}
\affiliation{Konkoly Observatory, MTA Research Centre for Astronomy and Earth Sciences, Konkoly Thege Mikl\'{o}s \'{u}t 15-17, H-1121 Budapest, Hungary}
%\label{firstpage}

\begin{abstract}
The search and study of close pairs of supermassive black holes (SMBHs) is important in the study of galaxy mergers which can possibly trigger active galactic nucleus (AGN) activity, and in the context of their evolution into the gravitational wave emitting regime. The quasar SDSS~J1048+0055 was identified as a SMBH binary candidate based on the observed double-peaked \OIII$\lambda\lambda$4959,5007 emission lines and two distinct radio components separated by $\sim 20$~pc \citep{2004ApJ...604L..33Z}. To ascertain the binary nature of this source, we analyzed multi-frequency, multi-epoch very long baseline interferometry (VLBI) data to investigate its pc-scale radio properties. The source shows double components with the western feature being brighter than the eastern one. This brighter %`core' 
component has a brightness temperature of  $\sim 10^{10}$~K, spectral index of $\alpha = -0.09 \pm 0.09$ (flat) and is indicative of mildly relativistic beaming. In contrast, the faint %`jet' 
component has a lower brightness temperature of $\sim 10^{8-9}$~K and steep spectrum. 
%These clues are consistent with a core--jet structure, as opposed to dual cores associated with a SMBH binary. Moreover, the apparent separation speed between the two components is much higher than the expected orbital motion in a binary SMBH. 
These clues are consistent with a core--jet structure, moreover, the apparent separation speed between the two components is much higher than the expected orbital motion in a binary SMBH. Thus the present study excludes the association of the two VLBI components with the cores of a SMBH binary, although the SMBH binary possibility (e.g., a pair of radio-loud and radio-quiet AGNs) is not fully ruled out.  
In the single active galactic nucleus (AGN) scenario, the double-peaked optical emission lines can originate from the jet interacting with the narrow-line region as indicated by a change in the jet direction at $\sim$ 140 pc.
\end{abstract}


%%%%%%%%%%%%%%%%%%%%%%%%%%%%%%%%%%%%%%%%%%%%%%%%%%%%%%%%%
\section{Introduction} \label{sec:intro}
%%%%%%%%%%%%%%%%%%%%%%%%%%%%%%%%%%%%%%%%%%%%%%%%%%%%%%%%%
A central supermassive black hole (SMBH, $\rm{M}_\bullet = 10^6-10^{10}~\mathrm{M}_\odot$) is putatively hosted by every massive galaxy \cite[e.g.][]{2005SSRv..116..523F,2013ARA&A..51..511K}. An efficient major merger in interacting galaxies can result in a gravitationally bound SMBH binary (SMBHB) with a typical separation of $\lesssim$ 10 pc \cite[e.g.][]{1980Natur.287..307B,2014SSRv..183..189C}. %A further merger of the SMBHB can drive the system into the gravitational wave emitting regime, detectable by upcoming/proposed space observatories \cite[e.g.][]{2013GWN.....6....4A}  or pulsar timing arrays \cite[e.g.][]{2010CQGra..27h4013H}. 
 The study of SMBHs can help understand their growth and role in triggering active galactic nucleus (AGN) activity in the merging system \citep[e.g.][]{2008ApJS..175..356H,2012ApJ...748L...7V,2018Natur.563..214K}. If one or both SMBHs of the binary system are actively accreting, the jet-based synchrotron radio emission can be used to trace the merger dynamics \cite[e.g.][]{2018RaSc...53.1211A,2018BSRSL..87..299D}.

A small sample of SMBHB and dual AGN (with kpc-scale separation) candidates have been reported to date \cite[e.g.][and references therein]{2018RaSc...53.1211A} and include AGN hosting a spatially resolved dual core morphology \citep[e.g.][]{2003ApJ...582L..15K,2006ApJ...646...49R,2011Natur.477..431F}. Methods of identification employ searches in spectroscopic \cite[double peaked spectral lines, e.g.][]{2009ApJ...705L..76W,2009ApJ...702L..82C,2010ApJ...715L..30L,2010ApJ...716..866S,2012ApJ...753...42C,2012ApJ...752...63S,2014MNRAS.437...32W} or timing domains \cite[periodic variability over year-timescales, e.g.][]{2015Natur.518...74G,2015Natur.525..351D,2016MNRAS.463.2145C}, or through imaged peculiar large-scale jet morphologies \cite[S, X or Z-shaped, e.g.][]{1978Natur.276..588E,2003ApJ...594L.103G,2014Natur.511...57D,2018Natur.563..214K}. Competing explanations can include AGN with overlapping spatially projected locations, outflows associated with the narrow line region \cite[e.g.][]{2010ApJ...716..131R,2013MNRAS.433.1161A}, or rotating gaseous disks \cite[e.g.][]{2005ApJ...627..721G} relevant to the spectroscopic domain; helical motion in the jet either due to large scale structured magnetic fields \cite[e.g.][]{2015ApJ...805...91M,2016MNRAS.463.1812M} or due to jet precession enabled by a misalignment between the jet direction and the accretion flow powering the central engine \cite[e.g.][]{2006ApJ...638..120C,2018MNRAS.474L..81L}, relevant to the timing and imaging domains. 

AGN hosting narrow optical spectral lines (e.g. [\ion{O}{iii}]) are key probes of ionized outflows likely originating in the narrow line region (NLR). The outflow kinematics and emission can be naturally powered by the central engine including the jet which can transfer momentum and energy flux, and the ionizing accretion disk and broad line region photons. Double peaked narrow lines are then indicative of a complex kinematic structuring in the NLR \cite[e.g.][]{2009ApJ...705L..20X} and can be used to probe mechanisms enabling it. Employing observational diagnostics including the optical line ratios provide constraints on the ionization parameter and distribution (density and geometry) of the NLR clouds which can then be used to infer their location and energetics \cite[e.g.][]{2005MNRAS.358.1043B,2006A&A...456..953B}. Radio very long baseline interferometry (VLBI) observations then help constrain the drivers of the ionized outflows and provide a consistent picture connecting the physical regions, complementing the observational diagnostics and hydrodynamic simulations of jet interaction with the interstellar medium \cite[e.g.][]{2012ApJ...757..136W}.

%In this paper, we present a systematic search of the binary quasar candidate in the object 
The quasar SDSS~J104807.74+005543.5 \citep[hereafter J1048+0055; redshift $z=0.6422$,][]{2004ApJ...604L..33Z} having RA=10$^\mathrm{h}$~48$^\mathrm{m}$~07\fs7446 and DEC=$+$00\deg~55$'$~43\farcs482\footnote{International Celestial Reference Frame 3, \url{http://hpiers.obspm.fr/icrs-pc/newwww/icrf/icrf3sx.txt}, P. Charlot et al., in prep.} is one of the first identified double-peaked narrow-line quasars found from the Sloan Digital Sky Survey Data Release~1 \citep[SDSS DR1,][]{2003AJ....126.2081A}. Its optical spectrum shows clear double peaks of the \OIII$\lambda\lambda$4959 and 5007 emission lines with a line luminosity corresponding to each peak comparable to that of a luminous quasar. As reported by \citet{2004ApJ...604L..33Z}, the observed wavelengths of the red and blue peaks for the \OIII$\lambda$4959 double-peaked line are $8159.5$ \ang~ (FWHM$=553$ \kms) and $8139.7$ \ang~ (FWHM$=610$ \kms) respectively, and that for the \OIII$\lambda$5007 double-peaked line are $8238.5 \pm 0.3$ \ang~ (FWHM$=553 \pm 25$ \kms) and $8218.5 \pm 0.5$ \ang~ (FWHM$=610 \pm 24$ \kms). The average velocity offsets are therefore $583$~\kms~ for the redshifted peaks and $-146$~\kms~ for the blueshifted peaks relative to the systemic redshift of J1048+0055. The \Hb line (observed wavelength$=7987.5 \pm 1.0$ \ang) is found to be broad with FWHM$=1238 \pm 102$ \kms, twice that of the \OIII lines, indicating that \Hb is blended with two narrow-line components.
The source shows two distinct radio features with a projected separation of $\sim 20$~pc in the 8.4 GHz image (resolution $\sim$ 1.5 mas  $\times$ 3 mas) from Very Long Baseline Array (VLBA) calibrator survey \citep[VCS1,][]{2002ApJS..141...13B}. \citet{2004ApJ...604L..33Z} speculate on the origin of the double-peaked \OIII and \Hb lines and suggest an origin from two distinct NLRs, with the identified radio components in 8.4 GHz VLBA image being two AGN cores. However, as the double-peaked optical emission lines from the nuclear region may also be produced by a range of distinguishable phenomena as mentioned above, the binary identification requires further supporting evidence. %Based on these evidences, they reported J1048+0055 as the possible binary quasar candidate. However, they leaved the space for the other possibilities too, such as the bipolar outflow. Our major motivation is to search for the presence or absence of the other features of binary quasars in this object.

In this paper, we employ archival multi-frequency, multi-epoch VLBI observations to investigate the validity of the proposed binary %nature 
identification of J1048+0055. %and speculate on the interesting connection between the central engine and the host galaxy through mechanical feedback involving powerful jet-driven outflows. 
%The basic properties of J1048+0055 are listed in Table~\ref{tab:prop}. 
Assuming a standard $\Lambda$CDM cosmology with $H_0 = 70$ \kms Mpc$^{-1}$, $\Omega_M = 0.3$ and $\Omega_V = 0.7$, an angular size of 1~mas corresponds to a linear size of 6.9~pc and a proper motion of 1~mas~yr$^{-1}$ to an apparent speed of 36.9~$c$ at the redshift of the source.

\begin{comment}
\begin{table*}
\centering
\caption{Basic properties of SDSS~J1048+0055}
\begin{tabular}{lll}
\hline
{\bf Parameters}         & {\bf Values}                                  & {\bf Reference}             \\
\hline
%Other name          & WISE J104807.74+005543.4                           & NED                         \\ 
RA                  & 10$^\mathrm{h}$ 48$^\mathrm{m}$ 07$^\mathrm{s}$.7446  & VLBA Calibrator Survey   \\
DEC                 & $+$00\deg~ 55$'$ 43$''$.482                           & VLBA Calibrator Survey   \\ 
%Object Type         & Quasar                                             & \citet{2003AJ....126.2579S} \\ 
Redshift            & $0.640855 \pm 0.000354$                            & \citet{2010MNRAS.405.2302H} \\ 
%Luminosity Distance & 3724 Mpc                                           & NED                         \\ 
Radio Loudness Parameter  & $(1.5 - 2.1)\times 10^4$                     & \citet{2004ApJ...604L..33Z} \\ 
%1.4 GHz Flux Density      & $271.5\pm 8.2$ mJy                   & NVSS, \citet{1998AJ....115.1693C}      \\
%Optical Absolute Magnitude ($i$-band)  & $-22.90\pm 0.51$     & SDSS DR2, \citet{2004AJ....128..502A}  \\ 
\hline
\end{tabular}
\label{tab:prop}
\end {table*}
\end{comment}

%%%%%%%%%%%%%%%%%%%%%%%%%%%%%%%%%%%%%%%%%%%%%%%%%%%%%%%%%
\section{Observations and data analysis}\label{sec:data}
%%%%%%%%%%%%%%%%%%%%%%%%%%%%%%%%%%%%%%%%%%%%%%%%%%%%%%%%%
The calibrated VLBA data at 2.3 and 8.4~GHz of J1048+0055 were obtained from the VLBA calibrator survey (VCS) archival repository\footnote{\url{http://astrogeo.org/cgi-bin/imdb_get_source.csh?source=J1048\%2B0055}} (VCS1: \citet{2002ApJS..141...13B} and VCS2: \citet{2016AJ....151..154G}), while the raw VLBA data at 1.4, 8.4 and 15.3~GHz of J1048+0055 from the National Radio Astronomy Observatory (NRAO) science data archive\footnote{\url{https://archive.nrao.edu/archive/advquery.jsp}}. The basic calibration of the raw data was performed using the Astronomical Image Processing System (AIPS) software package \citep{2003ASSL..285..109G}. While processing the raw data, we followed the standard procedure given in AIPS cookbook for VLBA data calibration \footnote{\url{ftp://ftp.aoc.nrao.edu/pub/software/aips/TEXT/PUBL/COOKC.PS.gz}}. The basic calibration process involves the data editing to flag essentially bad data, gain calibration for amplitude and phase corrections, fringe fitting to make the source at the phase center for improving phase accuracy, and bandpass calibration for correcting the change in frequency response across the channels. We further analyzed the calibrated datasets using the Caltech DIFMAP software package \citep{1995BAAS...27..903S}. Data with significant scatter are averaged in time over 30~seconds and bad data points are flagged. The edited data are then mapped and self-calibrated in a standard procedure. The final maps are made using uniform weighting to get higher resolution. The images are deconvolved with the beam using the CLEAN algorithm \citep{1974A&A....33..289H}. The CLEANing in the final maps is performed up to 3$\sigma$ level. The imaging parameters are given in Table~\ref{tab:image}.

\begin{deluxetable*}{cccccccc}[h!]
\tablecaption{Observing and imaging parameters\label{tab:image}}
\tabletypesize{\scriptsize}
\tablehead{
\colhead{Observing date} & \colhead{Project code} & \colhead{Frequency (Band)} & \colhead{Bandwidth} & \colhead{On-source time} & \colhead{Beam\tablenotemark{a}} & \colhead{RMS noise} & \colhead{Peak intensity}\\
\colhead{(yyyy-mm-dd)} & \colhead{} & \colhead{(GHz)} & \colhead{(MHz)} & \colhead{(min)} & \colhead{(mas$\times$mas, \degr)} & \colhead{(mJy~beam$^{-1}$)} & \colhead{(mJy~beam$^{-1}$)}
}
\startdata
 1995-07-15 & BB023  & 2.3 (S)  & 16  & 4.2 & $7.2\times 2.8, -5.7$   & 1.8 & 191 \\
 1995-07-15 & BB023  & 8.4 (X)  & 16  & 4.2 & $1.8\times 0.7, -3.8$   & 1.5 & 179 \\
 2003-03-21 & BP106  & 1.4 (L)  & 16  & 418 & $19.3\times 5.0, -20.1$ & 0.2 & 210 \\
 2005-07-25 & BL127  & 8.4 (X)  & 16  & 587 & $1.8\times 0.8, -5.9$   & 0.2 & 233 \\
 2005-07-25 & BL127  & 15.3 (U) & 16  & 569 & $1.0\times 0.5, -5.5$   & 0.3 & 219 \\
 2014-12-20 & BG219F & 2.3 (S)  & 128 & 2.7 & $6.8\times 2.7, -3.7$   & 0.6 & 261 \\
 2014-12-20 & BG219F & 8.4 (X)  & 384 & 2.7 & $1.8\times 0.7, -5.1$   & 0.3 & 168 \\ 
\enddata
\tablenotetext{a}{Restoring beam major and minor axes (FWHM) and the major axis position angle measured from north through east.}
\end{deluxetable*}

We performed circular Gaussian model fitting on the self-calibrated visibility data, and estimated the integrated flux densities of the two VLBI components (see Fig.~\ref{fig:images}), their angular size and relative positions of the faint component with respect to the bright one (see Table~\ref{tab:fitting}). We first fit a single Gaussian component on the source. However, multiple circular Gaussian components were used to fit the data when required to ensure that the model matches the data and there is no visible structure brighter than 5$\sigma$ appearing in the map. In cases of multi-component Gaussian models, the total flux density is obtained by adding that from individual components. With the use of elliptical Gaussian model fitting, the derived results are still consistent with those using circular Gaussian models. The errors in the model fitting parameters are estimated using the relations given in \citet{2012A&A...537A..70S} which are based on the approximations introduced by \citet{1999ASPC..180..301F} and modified for the strong side-lobes of VLBI observations. When the fitted Gaussian size $d$ of a component was found to be less than the minimum resolvable size $d_{\rm min}$ estimated using \citet{2005astro.ph..3225L} relation, the resolution-limited size $d_{\rm min}$ was used in place of $d$ as the size of that component.

\begin{deluxetable*}{cccccccccc}[h!]
\tablecaption{Fitting parameters\label{tab:fitting}}
\tabletypesize{\scriptsize}
\tablehead{
\colhead{Frequency band} & \multicolumn{3}{c|}{Flux density (mJy)} & \multicolumn{2}{c|}{Size ($\mu$as)} & \multicolumn{2}{l|}{Relative position of Jet comp. (mas)} & \multicolumn{2}{c}{Brightness temperature ($10^{10}$ K)}\\
\cline{2-10}
\colhead{(Observing year)} & Core comp. & jet comp. & Total  & Core comp. & jet comp. & $\Delta \alpha$ & $\Delta \delta$ & Core comp. & jet comp.
}
\startdata
 S (1995) & --- & --- & $273 \pm 30$ & $2028.0 \pm 22.4$ & --- & --- & --- & $2.58 \pm 0.29$ & ---\\
 X (1995) & $191 \pm 19$ & $13 \pm 6$ & $204 \pm 20$ & $261.0 \pm 2.6$ & $<764.9 \pm 88.6$ & $2.444 \pm 0.044$ & $-0.083 \pm 0.044$ & $8.09 \pm 0.84$ & $<0.06 \pm 0.03$\\
 L (2003) & --- & --- & $273 \pm 9$ & $2573.0 \pm 2.6$ & --- & --- & --- & $4.07 \pm 0.16$ & ---\\
 X (2005) & $252 \pm 8$ & $20 \pm 3$ & $272 \pm 8$ & $287.3 \pm 0.3$ & $648 \pm 10$ & $2.353 \pm 0.005$ & $-0.282 \pm 0.005$ & $8.64 \pm 0.27$ & $0.14 \pm 0.02$\\
 U (2005) & $239 \pm 11$ & $6 \pm 2$  & $245 \pm 11$ & $231.8 \pm 0.5$ & $<329.7 \pm 21.7$ & $2.294 \pm 0.011$ & $-0.212 \pm 0.011$ & $3.81 \pm 0.18$ & $<0.05 \pm 0.02$\\
 S (2014) & --- & --- & $358 \pm 32$ & $2125.0 \pm 15.3$ & --- & --- & --- & $3.02 \pm 0.29$ & ---\\
 X (2014) & $188 \pm 10$ & $14 \pm 3$ & $202 \pm 11$ & $300.1 \pm 0.9$ & $418.7 \pm 1.3$ & $2.224 \pm 0.001$ & $-0.295 \pm 0.001$ & $5.56 \pm 0.31$ & $0.21 \pm 0.05$\\  
\enddata
\end{deluxetable*}

%%%%%%%%%%%%%%%%%%%%%%%%%%%%%%%%%%%%%%%%%%%%%%%%%%%%%%%%%
\section{Results}\label{sec:results}
%%%%%%%%%%%%%%%%%%%%%%%%%%%%%%%%%%%%%%%%%%%%%%%%%%%%%%%%%

\subsection{Morphology}\label{sec:results:morpho}
\begin{figure}
\centering
\includegraphics[scale=0.4]{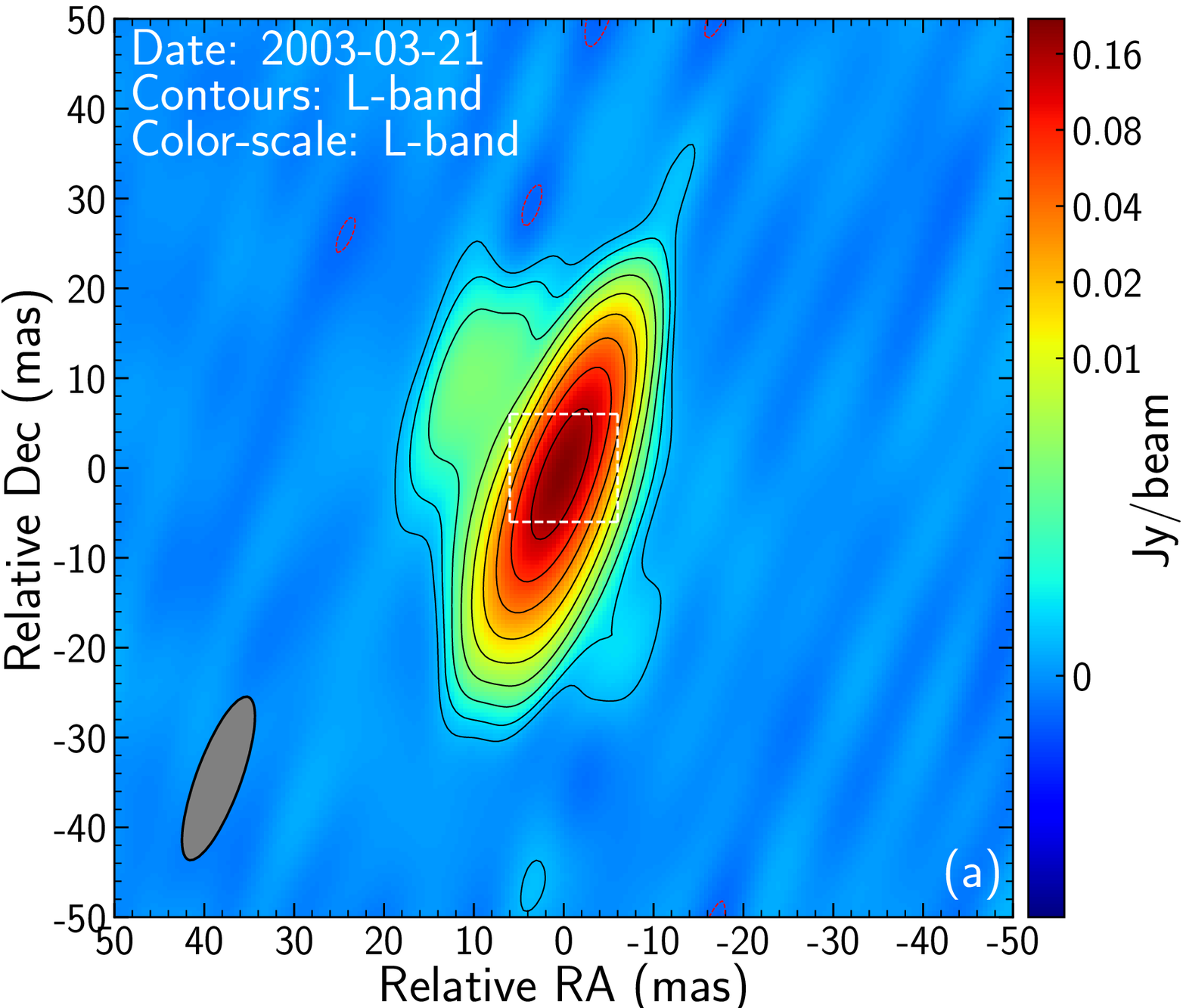}
\includegraphics[scale=0.4]{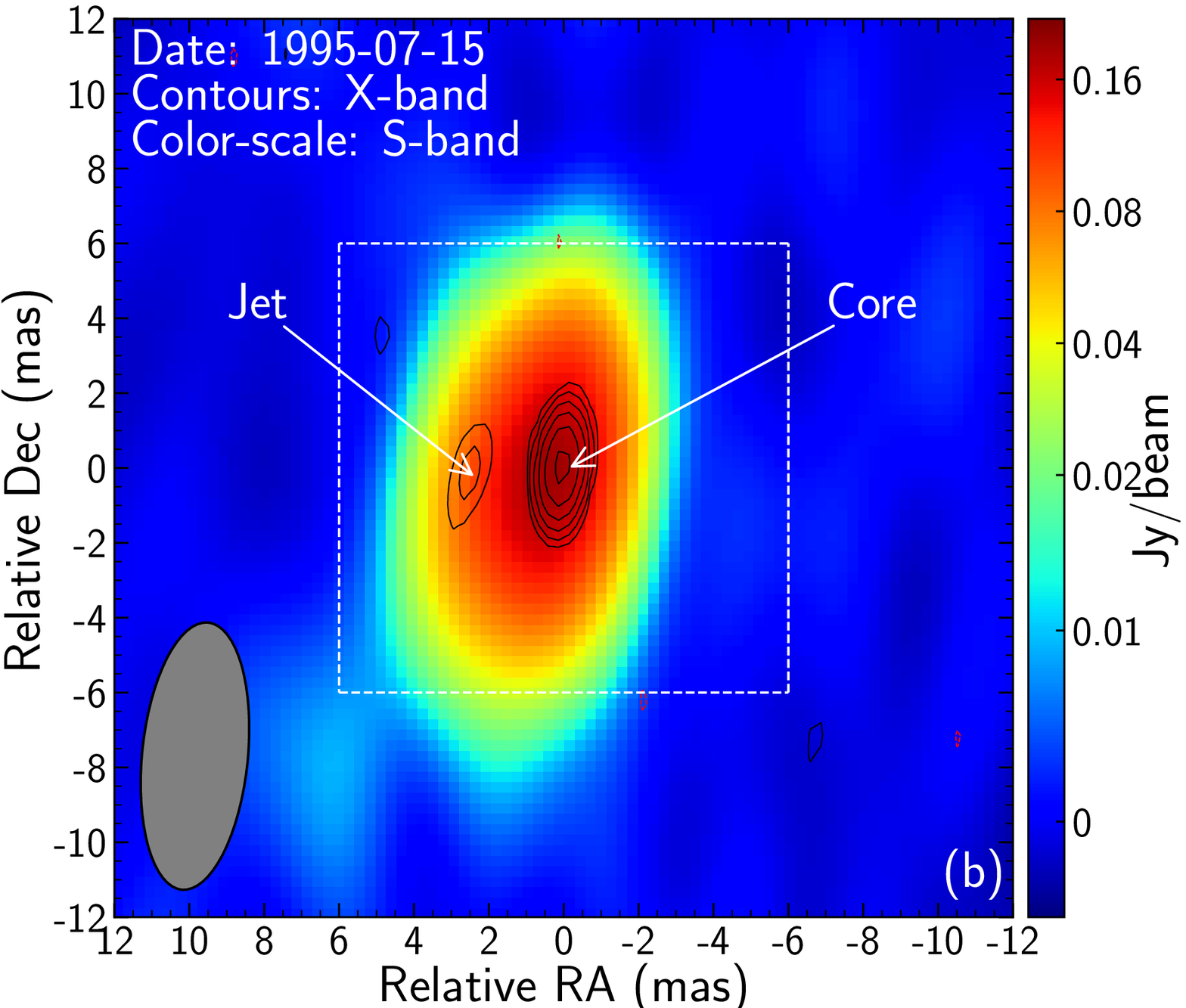}
\includegraphics[scale=0.4]{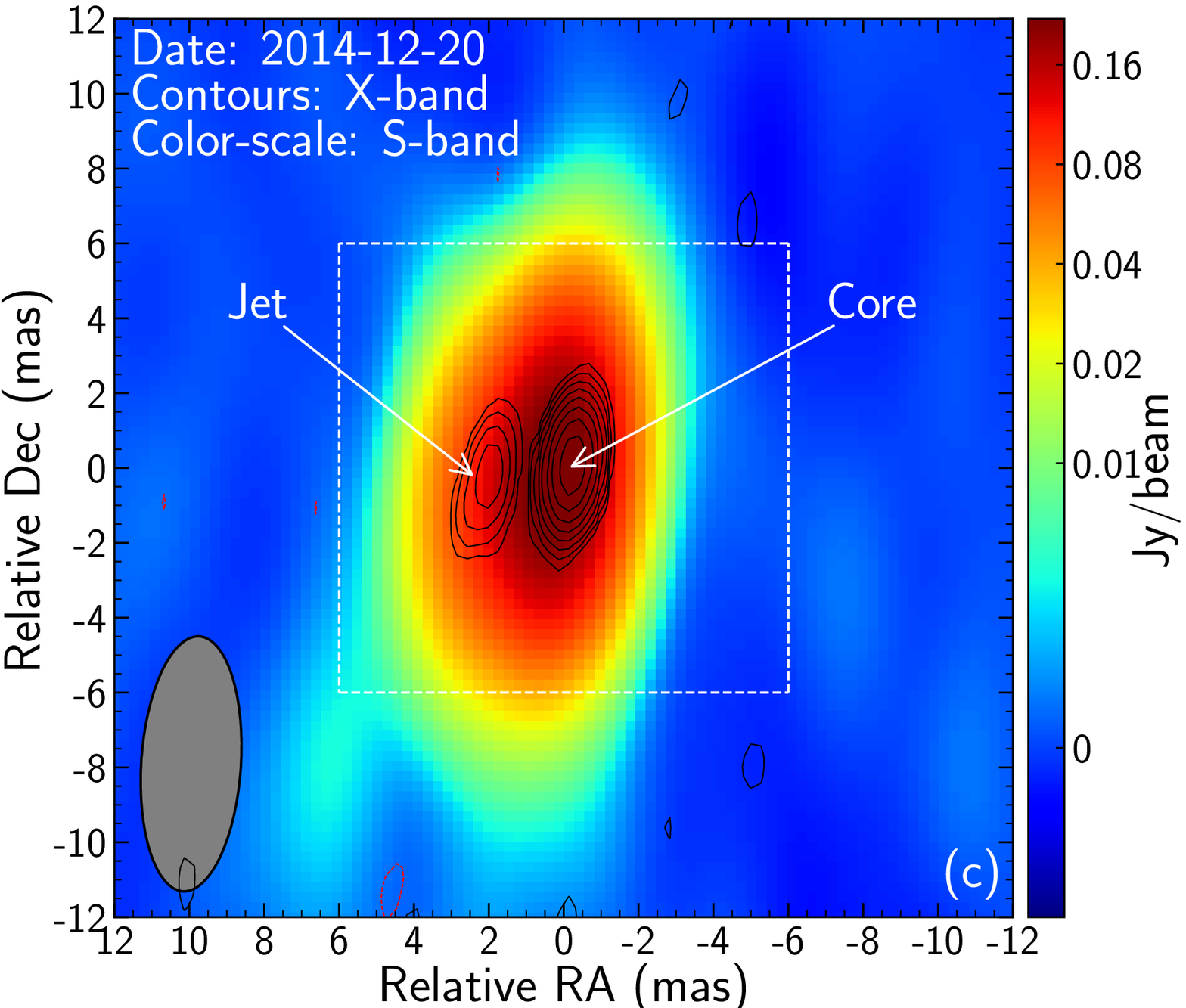}
\includegraphics[scale=0.4]{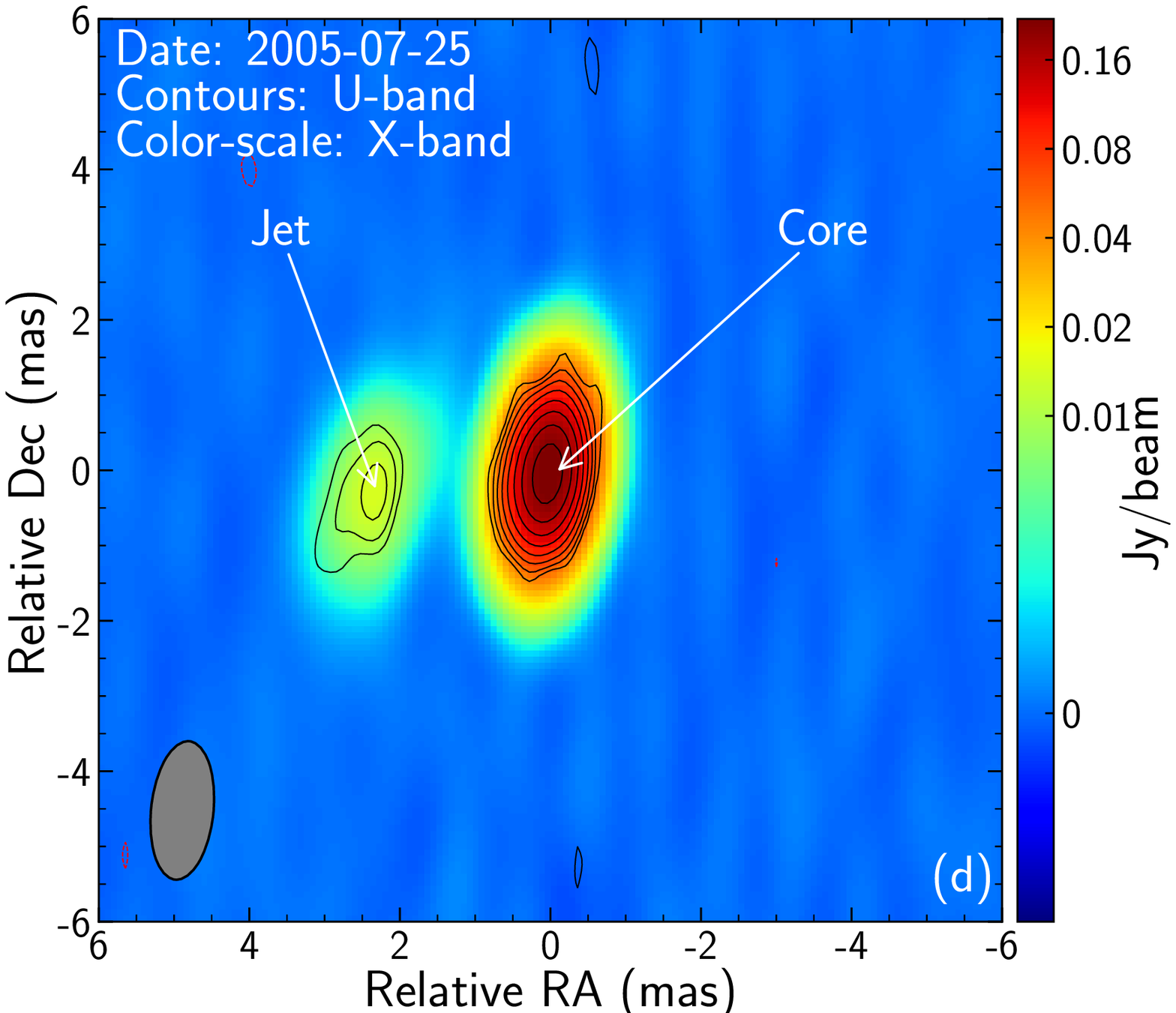}
\caption{\small The images of J1048+0055 in L, S, X and U bands at different epochs of observations. (a) The contours of L-band image at 2003 epoch overlaid on its color-scaled image. (b) The contours of X-band image at 1995 epoch overlaid on the color-scaled S-band image at 1995 epoch. (c) The contours of X-band image at 2014 epoch overlaid on the color-scaled S-band image at 2014 epoch. (d) The contours of U-band image at 2005 epoch overlaid on the color-scaled X-band image at 2005 epoch. The contour levels are given as $3\times rms \times (-1,1,2,4,8,16,32,64)$. The $rms$ in each image is referred in Table~\ref{tab:image}. The grey-colored ellipse in the bottom-left corner of each panel represents the restoring beam of the color-scaled image. The restoring beams for all the images are quantitatively given in Table~\ref{tab:image}. The color bar denotes the intensity of the map in logarithmic scale, in order to highlight the faint eastern component. The box shown by dashed line represents the region shown for the highest-resolution image in (d).}
\label{fig:images}
\end{figure}

The images of J1048+0055 for different wavebands and epochs of observation are shown in Figure~\ref{fig:images}. Two resolved components are seen in the X-band and U-band images. The western component is significantly brighter and more compact than the eastern one at both frequencies. The source is unresolved in the S-band images. However, the S-band images show an elongation in the East direction, in alignment with the resolved structure at two higher frequencies. 
%indicating the presence of the unresolved component. 
Also, the S-band image is fit better with an additional Gaussian component at the same position as that of the faint eastern component in X- and U-band images than a single component. The flux densities of the fitted components are given in Table~\ref{tab:fitting}. The L-band image shows a structure extended to about 20 mas in the northeast of the center, suggesting a jet deflection or bending based on a $\sim$ 40\deg~difference between the inner and outer jet direction. The north-east jet feature is real as it is detected at above 12$\sigma$ with a consequent smaller chance of being a sidelobe pattern. There is a hint of a counter-jet at around 3$\sigma$ which would require more sensitive imaging for a confirmation.﻿

\subsection{Spectral index}\label{sec:results:spex}
\begin{figure}
\centering
\includegraphics[scale=0.5]{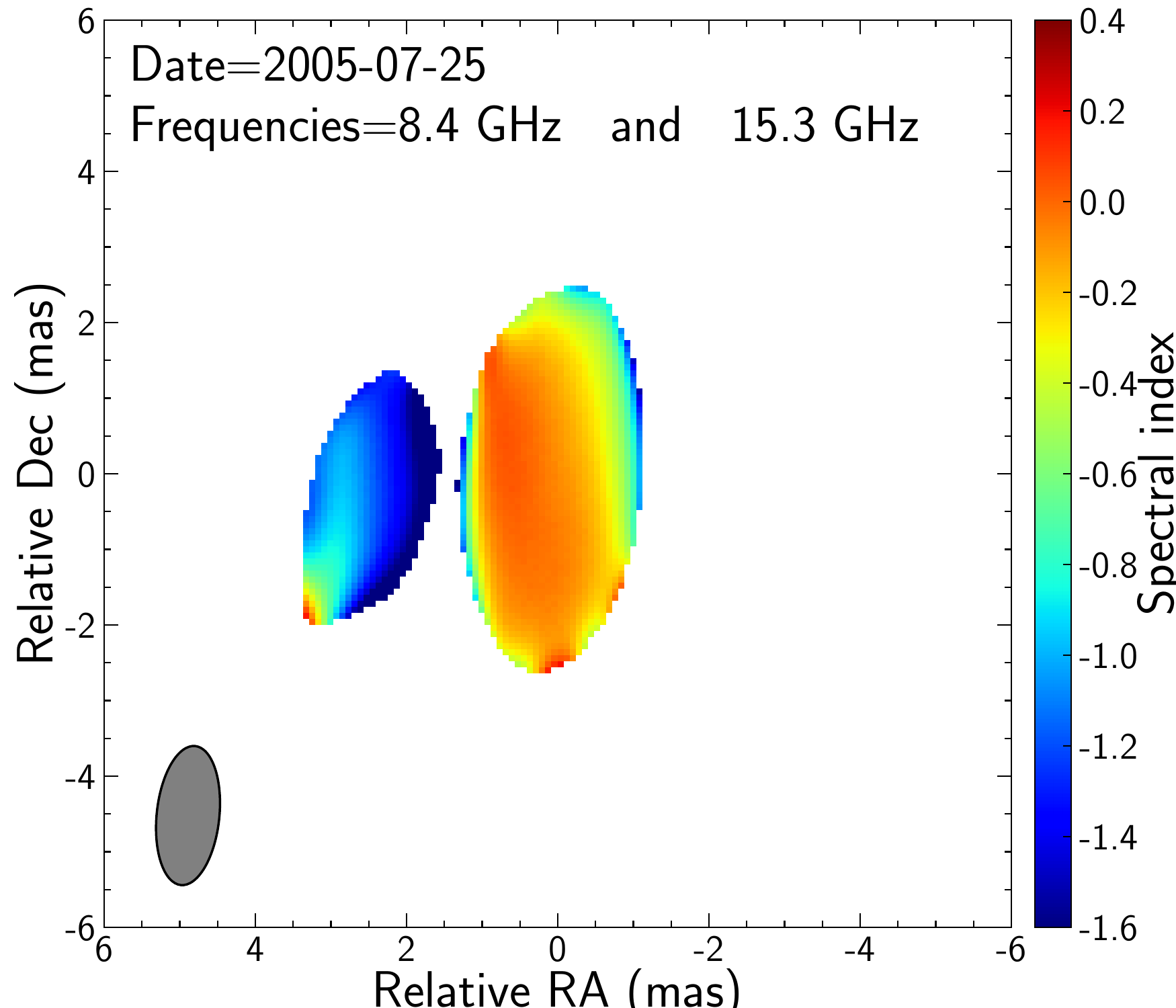}
\caption{\small Spectral index distribution from the X/U bands during 2005. %(a) The spectral index map of S- and X-band data in the 1995 epoch. (b) The spectral index map of X- and U-band data in the 2005 epoch. (c) The spectral index map of S- and X-band data in the 2014 epoch. 
The grey-colored ellipse in the bottom-left corner represents the restoring beam.}
\label{fig:SI}
\end{figure}

The spectral indices, $\alpha$  (defined as $S_\nu \propto \nu^{\alpha}$), are estimated for the epochs 1995, 2005 and 2014 using the model-fitted flux densities (see Table~\ref{tab:fitting}) of the contemporaneous dual frequency observations. The X/U band spectral index distribution during 2005 is also presented in Figure~\ref{fig:SI}. %While creating an spectral index map using the two different waveband images in the same epochs, 
We first aligned the two images by shifting their peak to the center of the images, then convolved the high-resolution U-band image with the restoring beam of the low-resolution X-band image, and changed the pixel scale of this convolved image equal to the X-band image, and finally made the spectral index map using the X-band image and the scaled U-band image after blanking the noise up to 3$\sigma$ level. %It can be seen for the 1995 and 2014 epochs that there is a negative gradient of spectral index variation towards East on the surface of the source, indicative of a transition of the emission property. However,
 The spectral index clearly indicates the flat spectrum with $\alpha = -0.09 \pm 0.09$ for the western (self-absorbed core, see more discussion in following subsections) component and steep spectrum with $\alpha = -2.02 \pm 0.18$ for the eastern (optically thin jet) component, and thereby provides direct evidence of the core--jet nature of the source J1048+0055. The average spectral indices of the overall source with unresolved components are $-0.22 \pm 0.08$ and $-0.43 \pm 0.04$ for the 1995 and 2014 data respectively. The difference in the spectral indices at 1995 and 2014 epochs could be related to the intrinsic change in the emission structure of the core region, as expected from the flux density changes listed in Table~\ref{tab:fitting}. However the resolution is not high enough to make a conclusive inference.

\subsection{Brightness temperature}\label{sec:results:temp}
The brightness temperature indicates the level of relativistically beamed jet emission; that for the AGN core is typically high ($\sim 10^{11} - 10^{12}$ K). The brightness temperatures $T_{\rm B}$ were estimated here (see Table~\ref{tab:fitting}) using \citep[e.g.,][]{1988gera.book..563K}:
\begin{equation}
\frac{T_{\rm B}}{1+z} = 1.22 \times 10^{12}~{\rm K} \left(\frac{S_{\rm total}}{\mathrm{Jy}}\right) \left(\frac{d}{\mathrm{mas}}\right)^{-2} \left(\frac{\nu}{\mathrm{GHz}}\right)^{-2},
\end{equation}
where $S_{\rm total}$ is the total flux density, $d$ the diameter (full width at half maximum, FWHM) of the source component, $\nu$ the observing frequency, and $z$ the redshift. The bright component indicates a $T_{\rm B}$ at least an order of magnitude higher than that of the faint component. The observed brightness temperatures of the bright component are close to but slightly lower than the equipartition brightness temperature \citep[$\sim 5 \times 10^{10}$~K,][]{1994ApJ...426...51R}, or the inverse Compton limit \citep[$\sim 10^{11}$~K,][]{1969ApJ...155L..71K}, suggesting that the radio jet in J1048+0055 is not strongly beamed. The $T_{\rm B}$ of the faint component is significantly (two orders of magnitude) lower than the typical $T_{\rm B}$ of Doppler-boosted radio-loud AGN cores. Instead, it is more consistent with a hot spot in the jet \citep[e.g. $10^{8-10}$~K in 3C~48,][]{2010MNRAS.402...87A}. Though, the distinctive differences in $T_{\rm B}$ (1--2 orders of magnitude) of the components, together with their spectral indices, confirm the core--jet nature. % (Sect.~\ref{sec:results:morpho}).

\subsection{Proper motion}\label{sec:results:promo}
As both components are detected and well separated in the X-band VLBI images in 1995, 2005 and 2014 (see Fig.~\ref{fig:images} and Table~\ref{tab:fitting}), the average proper motion of the jet with respect to the core is estimated to be $-0.013 \pm 0.001$ mas~year$^{-1}$ ($-0.49 \pm 0.03~c$). %, consistent with a source that is not strongly beamed. 
Any orbital motion of pc-scale separation in binary SMBHs cannot appear very fast. For instance, the VLBI observations of the archetypal compact SMBHB  0402+379 indicates a separation of 7.3~pc between two AGN cores \citep{2006ApJ...646...49R}, and a relative motion of $0.0054 \pm 0.0003 \, c$ is derived from the VLBI monitoring over 12 years \cite[][]{2017ApJ...843...14B}. The orbital velocity of a SMBHB (equating the binding energy to the kinetic energy of the system) is
\begin{equation}
v \sin i = 656~{\rm km~s^{-1}}\left(\frac{a}{{\rm pc}}\right)^{-1/2} \left((1+q)\frac{M_1}{10^8 M_\odot}\right)^{1/2},
\label{vorb}
\end{equation}
where $i$ is the inclination angle of the orbital plane with respect to the line of sight, $a$ is the binary separation, $q = M_2/M_1 \leq 1$ is the ratio of masses in the SMBHB system ($M_2$ and $M_1$ are the masses of the secondary and primary orbiting SMBHs, respectively). The SMBH in AGN can typically grow to masses of $\sim 10^{10} M_\odot$ \citep[e.g.][]{2008ApJ...680..169S,2016MNRAS.456L.109K}. The maximum orbital velocity for this system is therefore $v \lesssim 8 \times 10^{-3} ~ c$ for $q=1$ (SMBHB system with equal masses of $10^{10} M_\odot$), $a = 20$ pc and $i = 60$\deg~(mildly relativistic beaming inferred from the brightness temperatures suggests a moderate viewing angle between the jet direction and the observer line of sight). %{\bf The mildly relativistic beaming inferred from the brightness temperatures suggests a moderate viewing angle between the jet direction and the line of sight. Assuming the corresponding inclination angle $i=60$\deg, the orbital velocity will be $v \lesssim 8 \times 10^{-4} ~ c$ for $a = 20$ pc and an equal black hole mass $10^8 M_\odot$ for both primary and secondary.} 
%This corresponds to $v \lesssim 8 \times 10^{-4} ~ c$ for $q=1$ (SMBHB system with equal masses of $10^8 M_\odot$), $a = 20$ pc and $i = 60$\deg~(mildly relativistic beaming inferred from the brightness temperatures suggests a moderate viewing angle between the jet direction and the observer line of sight). 
The inferred relative motion $0.5~c$ in J1048+0055 is distinctly larger ($>$ two orders of magnitude) than expected from eqn.~(\ref{vorb}) thus adding to the evidence against the SMBHB scenario. 

\begin{comment}
\subsection{Variablity}\label{sec:results:vari}
The NRAO VLA Sky Survey (NVSS) in 1995 February \citep{1998AJ....115.1693C} and the Faint Images of the Radio Sky at Twenty centimeters (FIRST) survey in 1998 August \citep{1997ApJ...475..479W} at 1.4~GHz provide the flux-density estimates of the source J1048+0055 as $S_{\rm NVSS} = 271.5 \pm 8.2$~mJy and $S_{\rm FIRST} = 269.8 \pm 0.6$~mJy, which indicates a non-variable source. Note that our VLBI flux-density at 1.4~GHz during 2003 March is in agreement with these values. However, from Table~\ref{tab:fitting} we see that the VLBI flux density of J1048+0055 shows variablity in both S- and X-band. The S-band (2.3~GHz) flux density in the 2014 epoch is increased by $\sim$30\% compared to that during 1995. It should be noted that the S-band flux density is mainly contributed by the core, as indicated by the change of the peak brightness in the S-band images at the two epochs (see panels (b) and (c) in Figure~\ref{fig:images}). The X-band (8.4~GHz) flux density is found to be increasing in 2005 epoch from that in 1995 epoch and then decreasing in 2014 epoch. It is possible that the 8.4~GHz core flaring occurred at an earlier epoch compared to that in the 2.3~GHz core. Further dense VLBI monitoring observations are required in order to study this peculiar variability.
\end{comment}

%%%%%%%%%%%%%%%%%%%%%%%%%%%%%%%%%%%%%%%%%%%%%%%%%%%%%%%%%
\section{Re-examining the source nature}\label{sec:discuss}
%%%%%%%%%%%%%%%%%%%%%%%%%%%%%%%%%%%%%%%%%%%%%%%%%%%%%%%%%
The quasar J1048+0055 was reported to indicate double peaked \OIII$\lambda\lambda$4959,5007 emission lines and two distinct radio components at a projected separation of $\sim 20$ pc in VLBI images \citep{2004ApJ...604L..33Z}, prompting an investigation of its true nature with the availability of additional radio images. %The double-peaked emission lines in optical nuclear spectra of radio galaxies may be produced by binary SMBHs, jet-cloud interactions, bi-polar outflows, or rotating gaseous disk \citep[\textit{e.g.,}][]{2012ApJ...752...63S,2015ApJ...813..103M}. 

The steep spectrum, lower brightness temperature than expected in highly beamed AGN cores, and a larger proper motion speed compared to expected orbital motion of binary cores, as discussed in Section~\ref{sec:results}, rule out the identification of the eastern component as an additional self-absorbed core. The presence of double peaks in the optical spectrum and two distinct radio components are then not sufficient to identify the source as a SMBHB host, thus requiring a careful investigation from additional epochs of multi-wavelength observations to make a robust inference. 

The origin of the double peaked optical spectral lines presents an interesting avenue for interpretation. 
%The possibility of spatially projected distinct AGN may be ruled out on the basis of a non-detection of an additional self-absorbed core on transition from the arcsec-scale VLA 1.4 GHz to the mas-scale VLBA images. 
The compact emission structure, flat spectrum of the radio source and one-sided jet indicates a pole-on viewing direction. We assume this is the case for the following discussions. An asymmetry in the peak amplitudes of the optical emission lines \citep{2004ApJ...604L..33Z} renders an origin in an extended rotating gaseous disk with an underlying continuum source less feasible.
%{\bf The one-sided jet structure and flat spectrum of the radio source indicates a pole-on viewing direction. This makes the model of an extended rotating gaseous disk with an underlying continuum source less feasible. 
An origin in the NLR is then a likely scenario. %which interprets double-peaked narrow lines as a complex kinematic structuring in the NLR \cite[e.g.][]{2009ApJ...705L..20X}. the double peaks are produced by jet-cloud colliding, so another peak comes from counter-jet, but because of the beaming effect, the counter-jet is not visible in VLBI images. These double-peaked lines can be used to probe mechanisms enabling it. Employing observational diagnostics including the optical line ratios provide constraints on the ionization parameter and distribution (density and geometry) of the NLR clouds which can then be used to infer their location and energetics \cite[e.g.][]{2005MNRAS.358.1043B,2006A&A...456..953B}. Radio VLBI observations then help constrain the drivers of the ionized outflows and provide a consistent picture connecting the physical regions, complementing the observational diagnostics and hydrodynamic simulations of jet interaction with the interstellar medium \cite[e.g.][]{2012ApJ...757..136W}.}
The distance of the NLR clouds from the central engine can be approximated by assuming their photoionization by the underlying continuum emission source and is
\begin{equation}
r_{\rm NLR} = \left(\frac{Q}{4 \pi c n_{\rm H} U}\right)^{1/2} %= (145~{\rm pc})~\left(\frac{n_{\rm H}}{10^3~{\rm cm}^{-3}}\right)^(-1/2) \left(\frac{U}{10^{-2}}\right)^(-1/2),
\end{equation} 
where $n_{\rm H} \sim 10^3$ cm$^{-3}$ \cite[e.g.][]{2014MNRAS.440.3202V} is the number density of the hydrogen gas constituting a dense NLR, $U$ is the ionization parameter, and $Q \sim L/(h \nu) \sim 10^2 L_{\rm [OIII]}/(h \nu)$  \citep{1998ApJ...495..680B} is the rate at which ionizing photons emitted by the luminous quasar (with luminosity $L$) impinge on the NLR.
%$Q \sim 10^3 L_{\rm OIII}$ using which $Q = 3 \times 10^{45}$ erg s$^{-1}$ (0.23 L$_{\rm Eddington}$) 
 Employing $L_{\rm [OIII]} = 9.2 \times 10^{42}$~erg~s$^{-1}$, %and a $H_\alpha/H_\beta$ line ratio of $\sim$ 4 \citep{2004ApJ...604L..33Z}, 
$r_{\rm NLR} = 254$~pc is inferred, corresponding to an angular scale $\theta_{\rm NLR} \sim 37$~mas, close to the scales being probed in the L-band VLBA image (Fig.~\ref{fig:images} panel~(a)). This, the presence of an elongated outer jet component at $\sim$ 20~mas, and a difference in the direction of the inner and outer jet components suggest that the NLR kinematics may be driven by the jet. This could be enabled through radiative and momentum flux transport across a shock formed when the jet impacts the NLR cloud \cite[e.g.][]{1998ApJ...495..680B}, especially relevant to the production of the double peaked lines owing to a kinematic structuring upon impact \cite[e.g.][]{2015ApJ...799..161K}. %The available radio spectral power for the momentum flux transfer
%\begin{equation}
%P_{\nu} = 4 \pi D^2_L S_\nu (1+z)^{1-\alpha}.
%\end{equation}
The study of \cite{2008ApJ...686..859B} estimates an empirical relationship based on the best fitting linear relation between the observed total jet power and the spectral power from radio synchrotron emission (integrated over the full source: core and jet/lobes) from a sample of 18 sources, given by
\begin{equation}
\log \left(\frac{L}{10^{42} ~{\rm erg~s^{-1}}}\right) \sim 0.35 \log \left(\frac{P_{\rm 1.4}}{10^{24} ~{\rm W~Hz^{-1}}}\right)+1.85,
\end{equation}
where $P_{\rm 1.4} = P_{\nu} (\nu/{\rm 1.4~GHz})^{-\alpha}$ is the spectral power at 1.4 GHz corresponding to the spectral power for the momentum flux transfer $P_\nu = 4 \pi D^2_L S_\nu (1+z)^{1-\alpha}$. From the $S_\nu$ and $\alpha$ inferred in Table \ref{tab:fitting}, the average jet kinetic power $L \sim 6.8 \times 10^{45}$ erg s$^{-1}$. 

A jet-cloud interaction can cause a deflection \cite[e.g.][]{1998ApJ...495..680B,1998ApJ...502..199F,2005MNRAS.359..781S} which can be constrained and compared with observations to ascertain this scenario. The jet velocity \citep{1998ApJ...495..680B}
\begin{equation}
\beta = \frac{2 L}{c p_{\rm sh} A_{\rm sh}} f(\mathcal{M}_j,\Delta \theta),
\end{equation}
where $p_{\rm sh}$ is the pressure just behind the propagating shock acting on an area $A_{\rm sh}$ and $f(\mathcal{M}_j,\Delta \theta)$ is a factor parameterizing the dependence on the jet thermodynamic and kinematic properties through the Mach number $\mathcal{M}_j$ and the deflection angle $\Delta \theta$ respectively. Assuming that the jet with a half opening angle $\theta_0$, and proper motion speed $\beta \sim 0.5$, $L = 9.2 \times 10^{44} - 6.8 \times 10^{45}$ erg s$^{-1}$ acts through a shock dominated by ram pressure, $p_{\rm sh} = \rho_{\rm sh} v^2_{\rm sh} \sim m_{\rm H} n_{\rm H}~{\rm FWHM^2_{[OIII]}} = 6 \times 10^{-6}$ dyn cm$^{-2}$ with a surface area $A_{\rm sh} \sim r^2_{\rm NLR} \sin \theta_0 = 3.23 \times 10^{-2} \sin \theta_0$ kpc$^2$ (solid angle accounts for the area in contact with NLR clouds). Then, the factor $f(\mathcal{M}_j,\Delta \theta) \leq 1$ gives a $\theta_0$ in the range  $1.3^\circ - 9.9^\circ$, consistent with an observed median of $1.3^\circ$ from a large sample of AGN \citep{2017MNRAS.468.4992P}. With the parameterization \citep{1998ApJ...495..680B}
\begin{align}
&f(\mathcal{M}_j,\Delta \theta) = \left(\frac{1}{\gamma \mathcal{M}^2_j} + \frac{2}{\gamma+1} \left(\cos^2 \theta_1-\frac{1}{\mathcal{M}^2_j}\right)\right)\times\\ \nonumber
&\left(1+\frac{2}{\gamma-1} \frac{1}{\mathcal{M}^2_j}\right)^{-1} \sec \theta_1 \left(\sin (\theta_1+\Delta \theta)+\mathcal{M}_2  \cos (\theta_1+\Delta \theta)\right),
\end{align}
where $\gamma = 5/3$ is the adiabatic index, $\theta_1$ is the angle between the jet velocity and the normal of the shock front, and $\mathcal{M}_2 \sim 1/\sqrt{5}$ \citep{2002apa..book.....F} is the post-shock subsonic jet Mach number. Assuming a moderate to strong shock with $\mathcal{M}_j > 5$ and $\theta_1$ in the range 1$^\circ$ (near grazing) -- 90$^\circ$ (head-on), we obtain a maximum $f = 0.8$ corresponding to a deflection angle $\Delta \theta$ between $52^\circ - 76^\circ$ and an angle of approach $\theta_1 \leq$ 7.5\deg. The inferred $\Delta \theta$ is larger than the observed $\sim$ 40\deg, with the possibility of simplifying assumptions in the model including the structure and evolution of the interacting clouds and in the jet and shock energetics being a cause for this discrepancy. %{\bf This discrepancy is possibly because of the simplifying assumptions of photoionization mechanism, uniform and symmetric NLR clouds around the central engine, jet-driven NLR kinematics in the model.}
%, similar to the observed change of the projected jet position angle in Fig.~\ref{fig:images} (a). %A lower $\Delta \theta$ will correspond to a lower $f$ which can be better constrained with a more accurate observational measurement of $\theta_0$. 

%The VLBI images of  J1048+0055 clearly indicate a core--jet morphology characterizing the pc scale with less evidence for a SMBHB based on the results presented in the preceding section and supported by the above discussion. 
Based on the results presented in the preceding section and the above discussion, J1048+0055 indicates a core--jet morphology with less evidence for a SMBHB.
The blueshifted emission lines likely originate from the interaction between the jet and NLR clouds. The redshifted emission lines can then originate from the NLR region on the counter-jet side. The counter-jet component is however undetected in the VLBI images, likely due to Doppler de-boosted emission rendering it very faint. %, also possibly inferred from the asymmetry in profiles with the redshifted lines at lower fluxes. 
%{\bf The observed lower fluxes of redshifted emission lines is possibly caused by the very faint counter-jet component.}
To further address the origin of the double peaked emission lines, a search was conducted in a larger sky zone around J1048+0055 to find a second radio core at arcsecond separations. This resulted in a failure in such a detection up to a radius of $10\arcsec$ around J1048+0055 based on the NASA/IPAC Extragalactic Database (NED). We also looked at the Multi-Element Radio Linked Interferometer Network (MERLIN) L-band image of J1048+0055 and found the source remains unresolved at $\sim$200 mas ($\sim$1 kpc) resolution. This then rules out the presence of a dual radio-loud AGN at the kpc scale. The current analysis then tends to support a single radio-loud AGN, although the possibility of a second radio-quiet nucleus (e.g., a low-luminosity AGN) can not be fully excluded.

%%%%%%%%%%%%%%%%%%%%%%%%%%%%%%%%%%%%%%%%%%%%%%%%%%%%%%%%%
\section{Summary}\label{sec:summary}
%%%%%%%%%%%%%%%%%%%%%%%%%%%%%%%%%%%%%%%%%%%%%%%%%%%%%%%%%
The process of hierarchical merging should make close pairs of SMBHs common in the Universe, especially near the centers of galaxy clusters where multiple mergers have occurred. However, only a few candidates at sub-kpc separations have been seen to date. If they are radio-emitting AGN, their direct imaging and confirmation requires mas or even sub-mas resolution VLBI images at different wavebands and at well-separated epochs of observation. The present systematic VLBI imaging of the quasar J1048$+$0055 produces the following inferences.
\begin{enumerate}
\item[1.] Identification of a core--jet morphology from significant differences in measured flux densities, spectral index (flat--steep) and brightness temperature in the resolved components and a jet separation speed of $\sim 0.5~c$, unlikely to be the case for a SMBHB system at a separation of 20 pc. 
\item[2.] Indication of a mildly relativistic jet from the brightness temperature (mean of $\sim 5 \times 10^{10}$ K) and proper motion. 
\item[3.] The double-peaked optical emission lines originating from the interaction of jet with the NLR cloud as indicated by a change in the jet direction at $\sim$ 138 pc (20 mas).
\end{enumerate}
Though J1048+0055 may not host a SMBHB system, it is a promising target to study jet--NLR interaction, the origin of lower velocity outflows ($\sim$ 1000 km s$^{-1}$) and the structure of their host regions. Additional evidence can be obtained by the inference of polarized emission at these scales owing to the jet interaction with the NLR clouds as it can cause pockets with an increased plasma density.

%%%%%%%%%%%%%%%%%%%%%%%%%%%%%%%%%%%%%%%%%%%%%%%%%%%%%%%%%
\acknowledgements
%%%%%%%%%%%%%%%%%%%%%%%%%%%%%%%%%%%%%%%%%%%%%%%%%%%%%%%%%
We would like to thank the referee for his/her valuable comments and suggestions which have improved our manuscript. This work was supported by National Key R\&D Programme of China (2018YFA0404603). PM is supported by the NSFC Research Fund for International Young Scientists (grant no. 11650110438). 
%This research has made use of NASA's Astrophysics Data System.
SF thanks for the support received from the Hungarian Research, Development and Innovation Office (OTKA NN110333). SJ thanks Yingkang Zhang and Xiaofeng Li for helping in VLBI data analysis. The National Radio Astronomy Observatory is a facility of the National Science Foundation operated under cooperative agreement by Associated Universities, Inc. We are grateful to Yuri Y. Kovalev and Alexander Pushkarev for making their fully calibrated VLBI FITS data publicly available and to Leonid Petrov for maintaining, at the Astrogeo Center, the data base of brightness distributions, correlated flux densities and images of compact radio sources produced with VLBI. MERLIN is a National Facility operated by the University of Manchester at Jodrell Bank Observatory on behalf of STFC. The NASA/IPAC Extragalactic Database (NED) is operated by the Jet Propulsion Laboratory, California Institute of Technology, under contract with the National Aeronautics and Space Administration.

%%%%%%%%%%%%%%%%%%%%%%%%%%%%%%%%%%%%%%%%%%%%%%%%%%%%%%%%%
\bibliography{references}

\begin{thebibliography}{}
\expandafter\ifx\csname natexlab\endcsname\relax\def\natexlab#1{#1}\fi
\providecommand{\url}[1]{\href{#1}{#1}}
\providecommand{\dodoi}[1]{doi:~\href{http://doi.org/#1}{\nolinkurl{#1}}}
\providecommand{\doeprint}[1]{\href{http://ascl.net/#1}{\nolinkurl{http://ascl.net/#1}}}
\providecommand{\doarXiv}[1]{\href{https://arxiv.org/abs/#1}{\nolinkurl{https://arxiv.org/abs/#1}}}

\bibitem[{{Abazajian} {et~al.}(2003){Abazajian}, {Adelman-McCarthy},
  {Ag{\"u}eros}, {Allam}, {Anderson}, {Annis}, {Bahcall}, {Baldry}, {Bastian},
  {Berlind}, {Bernardi}, {Blanton}, {Blythe}, \&
  {Bochanski}}]{2003AJ....126.2081A}
{Abazajian}, K., {Adelman-McCarthy}, J.~K., {Ag{\"u}eros}, M.~A., {et~al.}
  2003, \aj, 126, 2081, \dodoi{10.1086/378165}

\bibitem[{{An} {et~al.}(2018){An}, {Mohan}, \& {Frey}}]{2018RaSc...53.1211A}
{An}, T., {Mohan}, P., \& {Frey}, S. 2018, Radio Science, 53, 1211,
  \dodoi{10.1029/2018RS006647}

\bibitem[{{An} {et~al.}(2010){An}, {Hong}, {Hardcastle}, {Worrall}, {Venturi},
  {Pearson}, {Shen}, {Zhao}, \& {Feng}}]{2010MNRAS.402...87A}
{An}, T., {Hong}, X.~Y., {Hardcastle}, M.~J., {et~al.} 2010, \mnras, 402, 87,
  \dodoi{10.1111/j.1365-2966.2009.15899.x}

\bibitem[{{An} {et~al.}(2013){An}, {Paragi}, {Frey}, {Xiao}, {Baan}, {Komossa},
  {Gab{\'a}nyi}, {Xu}, \& {Hong}}]{2013MNRAS.433.1161A}
{An}, T., {Paragi}, Z., {Frey}, S., {et~al.} 2013, \mnras, 433, 1161,
  \dodoi{10.1093/mnras/stt801}

\bibitem[{{Bansal} {et~al.}(2017){Bansal}, {Taylor}, {Peck}, {Zavala}, \&
  {Romani}}]{2017ApJ...843...14B}
{Bansal}, K., {Taylor}, G.~B., {Peck}, A.~B., {Zavala}, R.~T., \& {Romani},
  R.~W. 2017, \apj, 843, 14, \dodoi{10.3847/1538-4357/aa74e1}

\bibitem[{{Baskin} \& {Laor}(2005)}]{2005MNRAS.358.1043B}
{Baskin}, A., \& {Laor}, A. 2005, \mnras, 358, 1043,
  \dodoi{10.1111/j.1365-2966.2005.08841.x}

\bibitem[{{Beasley} {et~al.}(2002){Beasley}, {Gordon}, {Peck}, {Petrov},
  {MacMillan}, {Fomalont}, \& {Ma}}]{2002ApJS..141...13B}
{Beasley}, A.~J., {Gordon}, D., {Peck}, A.~B., {et~al.} 2002, \apjs, 141, 13,
  \dodoi{10.1086/339806}

\bibitem[{{Begelman} {et~al.}(1980){Begelman}, {Blandford}, \&
  {Rees}}]{1980Natur.287..307B}
{Begelman}, M.~C., {Blandford}, R.~D., \& {Rees}, M.~J. 1980, \nat, 287, 307,
  \dodoi{10.1038/287307a0}

\bibitem[{{Bennert} {et~al.}(2006){Bennert}, {Jungwiert}, {Komossa}, {Haas}, \&
  {Chini}}]{2006A&A...456..953B}
{Bennert}, N., {Jungwiert}, B., {Komossa}, S., {Haas}, M., \& {Chini}, R. 2006,
  \aap, 456, 953, \dodoi{10.1051/0004-6361:20065319}

\bibitem[{{Bicknell} {et~al.}(1998){Bicknell}, {Dopita}, {Tsvetanov}, \&
  {Sutherland}}]{1998ApJ...495..680B}
{Bicknell}, G.~V., {Dopita}, M.~A., {Tsvetanov}, Z.~I., \& {Sutherland}, R.~S.
  1998, \apj, 495, 680, \dodoi{10.1086/305336}

\bibitem[{{B{\^i}rzan} {et~al.}(2008){B{\^i}rzan}, {McNamara}, {Nulsen},
  {Carilli}, \& {Wise}}]{2008ApJ...686..859B}
{B{\^i}rzan}, L., {McNamara}, B.~R., {Nulsen}, P.~E.~J., {Carilli}, C.~L., \&
  {Wise}, M.~W. 2008, \apj, 686, 859, \dodoi{10.1086/591416}

\bibitem[{{Caproni} {et~al.}(2006){Caproni}, {Abraham}, \& {Mosquera
  Cuesta}}]{2006ApJ...638..120C}
{Caproni}, A., {Abraham}, Z., \& {Mosquera Cuesta}, H.~J. 2006, \apj, 638, 120,
  \dodoi{10.1086/498684}

\bibitem[{{Charisi} {et~al.}(2016){Charisi}, {Bartos}, {Haiman},
  {Price-Whelan}, {Graham}, {Bellm}, {Laher}, \&
  {M{\'a}rka}}]{2016MNRAS.463.2145C}
{Charisi}, M., {Bartos}, I., {Haiman}, Z., {et~al.} 2016, \mnras, 463, 2145,
  \dodoi{10.1093/mnras/stw1838}

\bibitem[{{Colpi}(2014)}]{2014SSRv..183..189C}
{Colpi}, M. 2014, Space Sci. Rev., 183, 189, \dodoi{10.1007/s11214-014-0067-1}

\bibitem[{{Comerford} {et~al.}(2012){Comerford}, {Gerke}, {Stern}, {Cooper},
  {Weiner}, {Newman}, {Madsen}, \& {Barrows}}]{2012ApJ...753...42C}
{Comerford}, J.~M., {Gerke}, B.~F., {Stern}, D., {et~al.} 2012, \apj, 753, 42,
  \dodoi{10.1088/0004-637X/753/1/42}

\bibitem[{{Comerford} {et~al.}(2009){Comerford}, {Griffith}, {Gerke}, {Cooper},
  {Newman}, {Davis}, \& {Stern}}]{2009ApJ...702L..82C}
{Comerford}, J.~M., {Griffith}, R.~L., {Gerke}, B.~F., {et~al.} 2009, \apjl,
  702, L82, \dodoi{10.1088/0004-637X/702/1/L82}

\bibitem[{{Das} {et~al.}(2018){Das}, {Rubinur}, {Kharb}, {Varghese},
  {Novakkuni}, \& {James}}]{2018BSRSL..87..299D}
{Das}, M., {Rubinur}, K., {Kharb}, P., {et~al.} 2018, Bulletin de la
  Soci\'{e}t\'{e} Royale des Sciences de Li\`{e}ge, 87, 299.
\newblock \doarXiv{1708.01185}

\bibitem[{{Deane} {et~al.}(2014){Deane}, {Paragi}, {Jarvis}, {Coriat},
  {Bernardi}, {Fender}, {Frey}, {Heywood}, {Kl{\"o}ckner}, {Grainge}, \&
  {Rumsey}}]{2014Natur.511...57D}
{Deane}, R.~P., {Paragi}, Z., {Jarvis}, M.~J., {et~al.} 2014, \nat, 511, 57,
  \dodoi{10.1038/nature13454}

\bibitem[{{D'Orazio} {et~al.}(2015){D'Orazio}, {Haiman}, \&
  {Schiminovich}}]{2015Natur.525..351D}
{D'Orazio}, D.~J., {Haiman}, Z., \& {Schiminovich}, D. 2015, \nat, 525, 351,
  \dodoi{10.1038/nature15262}

\bibitem[{{Ekers} {et~al.}(1978){Ekers}, {Fanti}, {Lari}, \&
  {Parma}}]{1978Natur.276..588E}
{Ekers}, R.~D., {Fanti}, R., {Lari}, C., \& {Parma}, P. 1978, \nat, 276, 588,
  \dodoi{10.1038/276588a0}

\bibitem[{{Fabbiano} {et~al.}(2011){Fabbiano}, {Wang}, {Elvis}, \&
  {Risaliti}}]{2011Natur.477..431F}
{Fabbiano}, G., {Wang}, J., {Elvis}, M., \& {Risaliti}, G. 2011, \nat, 477,
  431, \dodoi{10.1038/nature10364}

\bibitem[{{Falcke} {et~al.}(1998){Falcke}, {Wilson}, \&
  {Simpson}}]{1998ApJ...502..199F}
{Falcke}, H., {Wilson}, A.~S., \& {Simpson}, C. 1998, \apj, 502, 199,
  \dodoi{10.1086/305886}

\bibitem[{{Ferrarese} \& {Ford}(2005)}]{2005SSRv..116..523F}
{Ferrarese}, L., \& {Ford}, H. 2005, Space Sci. Rev., 116, 523,
  \dodoi{10.1007/s11214-005-3947-6}

\bibitem[{{Fomalont}(1999)}]{1999ASPC..180..301F}
{Fomalont}, E.~B. 1999, in Astronomical Society of the Pacific Conference
  Series, Vol. 180, Synthesis Imaging in Radio Astronomy II, ed. G.~B.
  {Taylor}, C.~L. {Carilli}, \& R.~A. {Perley}, 301

\bibitem[{{Frank} {et~al.}(2002){Frank}, {King}, \&
  {Raine}}]{2002apa..book.....F}
{Frank}, J., {King}, A., \& {Raine}, D.~J. 2002, {Accretion Power in
  Astrophysics: Third Edition} (Cambridge University Press, Cambridge), 398

\bibitem[{{Gopal-Krishna} {et~al.}(2003){Gopal-Krishna}, {Biermann}, \&
  {Wiita}}]{2003ApJ...594L.103G}
{Gopal-Krishna}, {Biermann}, P.~L., \& {Wiita}, P.~J. 2003, \apjl, 594, L103,
  \dodoi{10.1086/378766}

\bibitem[{{Gordon} {et~al.}(2016){Gordon}, {Jacobs}, {Beasley}, {Peck},
  {Gaume}, {Charlot}, {Fey}, {Ma}, {Titov}, \& {Boboltz}}]{2016AJ....151..154G}
{Gordon}, D., {Jacobs}, C., {Beasley}, A., {et~al.} 2016, \aj, 151, 154,
  \dodoi{10.3847/0004-6256/151/6/154}

\bibitem[{{Graham} {et~al.}(2015){Graham}, {Djorgovski}, {Stern}, {Glikman},
  {Drake}, {Mahabal}, {Donalek}, {Larson}, \&
  {Christensen}}]{2015Natur.518...74G}
{Graham}, M.~J., {Djorgovski}, S.~G., {Stern}, D., {et~al.} 2015, \nat, 518,
  74, \dodoi{10.1038/nature14143}

\bibitem[{{Greene} \& {Ho}(2005)}]{2005ApJ...627..721G}
{Greene}, J.~E., \& {Ho}, L.~C. 2005, \apj, 627, 721, \dodoi{10.1086/430590}

\bibitem[{{Greisen}(2003)}]{2003ASSL..285..109G}
{Greisen}, E.~W. 2003, in Astrophysics and Space Science Library, Vol. 285,
  Information Handling in Astronomy - Historical Vistas, ed. A.~{Heck}, 109

\bibitem[{{Hogbom} \& {Brouw}(1974)}]{1974A&A....33..289H}
{Hogbom}, J.~A., \& {Brouw}, W.~N. 1974, \aap, 33, 289

\bibitem[{{Hopkins} {et~al.}(2008){Hopkins}, {Hernquist}, {Cox}, \& {Kere{\v
  s}}}]{2008ApJS..175..356H}
{Hopkins}, P.~F., {Hernquist}, L., {Cox}, T.~J., \& {Kere{\v s}}, D. 2008,
  \apjs, 175, 356, \dodoi{10.1086/524362}

\bibitem[{{Kellermann} \& {Owen}(1988)}]{1988gera.book..563K}
{Kellermann}, K.~I., \& {Owen}, F.~N. 1988, {in {Kellermann}, K.~I. and
  {Verschuur}, G.~L., eds, Galactic and Extragalactic Radio Astronomy, 2nd
  edition}, ed. K.~I. {Kellermann} \& G.~L. {Verschuur}, Springer-Verlag
  (Springer-Verlag, Berlin--New York), 563--602

\bibitem[{{Kellermann} \& {Pauliny-Toth}(1969)}]{1969ApJ...155L..71K}
{Kellermann}, K.~I., \& {Pauliny-Toth}, I.~I.~K. 1969, \apjl, 155, L71,
  \dodoi{10.1086/180305}

\bibitem[{{Kharb} {et~al.}(2015){Kharb}, {Das}, {Paragi}, {Subramanian}, \&
  {Chitta}}]{2015ApJ...799..161K}
{Kharb}, P., {Das}, M., {Paragi}, Z., {Subramanian}, S., \& {Chitta}, L.~P.
  2015, \apj, 799, 161, \dodoi{10.1088/0004-637X/799/2/161}

\bibitem[{{King}(2016)}]{2016MNRAS.456L.109K}
{King}, A. 2016, \mnras, 456, L109, \dodoi{10.1093/mnrasl/slv186}

\bibitem[{{Komossa} {et~al.}(2003){Komossa}, {Burwitz}, {Hasinger}, {Predehl},
  {Kaastra}, \& {Ikebe}}]{2003ApJ...582L..15K}
{Komossa}, S., {Burwitz}, V., {Hasinger}, G., {et~al.} 2003, \apjl, 582, L15,
  \dodoi{10.1086/346145}

\bibitem[{{Kormendy} \& {Ho}(2013)}]{2013ARA&A..51..511K}
{Kormendy}, J., \& {Ho}, L.~C. 2013, \araa, 51, 511,
  \dodoi{10.1146/annurev-astro-082708-101811}

\bibitem[{{Koss} {et~al.}(2018){Koss}, {Blecha}, {Bernhard}, {Hung}, {Lu},
  {Trakthenbrot}, {Treister}, {Weigel}, {Sartori}, {Mushotzky}, {Schawinski},
  {Ricci}, {Veilleux}, \& {Sanders}}]{2018Natur.563..214K}
{Koss}, M.~J., {Blecha}, L., {Bernhard}, P., {et~al.} 2018, \nat, 563, 214,
  \dodoi{10.1038/s41586-018-0652-7}

\bibitem[{{Liska} {et~al.}(2018){Liska}, {Hesp}, {Tchekhovskoy}, {Ingram}, {van
  der Klis}, \& {Markoff}}]{2018MNRAS.474L..81L}
{Liska}, M., {Hesp}, C., {Tchekhovskoy}, A., {et~al.} 2018, \mnras, 474, L81,
  \dodoi{10.1093/mnrasl/slx174}

\bibitem[{{Liu} {et~al.}(2010){Liu}, {Greene}, {Shen}, \&
  {Strauss}}]{2010ApJ...715L..30L}
{Liu}, X., {Greene}, J.~E., {Shen}, Y., \& {Strauss}, M.~A. 2010, \apjl, 715,
  L30, \dodoi{10.1088/2041-8205/715/1/L30}

\bibitem[{{Lobanov}(2005)}]{2005astro.ph..3225L}
{Lobanov}, A.~P. 2005, arXiv:astro-ph/050322, \dodoi{arXiv:astro-ph/0503225}

\bibitem[{{Mohan} {et~al.}(2016){Mohan}, {An}, {Frey}, {Mangalam},
  {Gab{\'a}nyi}, \& {Kun}}]{2016MNRAS.463.1812M}
{Mohan}, P., {An}, T., {Frey}, S., {et~al.} 2016, \mnras, 463, 1812,
  \dodoi{10.1093/mnras/stw2154}

\bibitem[{{Mohan} \& {Mangalam}(2015)}]{2015ApJ...805...91M}
{Mohan}, P., \& {Mangalam}, A. 2015, \apj, 805, 91,
  \dodoi{10.1088/0004-637X/805/2/91}

\bibitem[{{Pushkarev} {et~al.}(2017){Pushkarev}, {Kovalev}, {Lister}, \&
  {Savolainen}}]{2017MNRAS.468.4992P}
{Pushkarev}, A.~B., {Kovalev}, Y.~Y., {Lister}, M.~L., \& {Savolainen}, T.
  2017, \mnras, 468, 4992, \dodoi{10.1093/mnras/stx854}

\bibitem[{{Readhead}(1994)}]{1994ApJ...426...51R}
{Readhead}, A.~C.~S. 1994, \apj, 426, 51, \dodoi{10.1086/174038}

\bibitem[{{Rodriguez} {et~al.}(2006){Rodriguez}, {Taylor}, {Zavala}, {Peck},
  {Pollack}, \& {Romani}}]{2006ApJ...646...49R}
{Rodriguez}, C., {Taylor}, G.~B., {Zavala}, R.~T., {et~al.} 2006, \apj, 646,
  49, \dodoi{10.1086/504825}

\bibitem[{{Rosario} {et~al.}(2010){Rosario}, {Shields}, {Taylor}, {Salviander},
  \& {Smith}}]{2010ApJ...716..131R}
{Rosario}, D.~J., {Shields}, G.~A., {Taylor}, G.~B., {Salviander}, S., \&
  {Smith}, K.~L. 2010, \apj, 716, 131, \dodoi{10.1088/0004-637X/716/1/131}

\bibitem[{{Saxton} {et~al.}(2005){Saxton}, {Bicknell}, {Sutherland}, \&
  {Midgley}}]{2005MNRAS.359..781S}
{Saxton}, C.~J., {Bicknell}, G.~V., {Sutherland}, R.~S., \& {Midgley}, S. 2005,
  \mnras, 359, 781, \dodoi{10.1111/j.1365-2966.2005.08962.x}

\bibitem[{{Schinzel} {et~al.}(2012){Schinzel}, {Lobanov}, {Taylor}, {Jorstad},
  {Marscher}, \& {Zensus}}]{2012A&A...537A..70S}
{Schinzel}, F.~K., {Lobanov}, A.~P., {Taylor}, G.~B., {et~al.} 2012, \aap, 537,
  A70, \dodoi{10.1051/0004-6361/201117705}

\bibitem[{{Shen} {et~al.}(2008){Shen}, {Greene}, {Strauss}, {Richards}, \&
  {Schneider}}]{2008ApJ...680..169S}
{Shen}, Y., {Greene}, J.~E., {Strauss}, M.~A., {Richards}, G.~T., \&
  {Schneider}, D.~P. 2008, \apj, 680, 169, \dodoi{10.1086/587475}

\bibitem[{{Shepherd} {et~al.}(1995){Shepherd}, {Pearson}, \&
  {Taylor}}]{1995BAAS...27..903S}
{Shepherd}, M.~C., {Pearson}, T.~J., \& {Taylor}, G.~B. 1995, in BAAS, Vol. 27,
  p. 903

\bibitem[{{Smith} {et~al.}(2010){Smith}, {Shields}, {Bonning}, {McMullen},
  {Rosario}, \& {Salviander}}]{2010ApJ...716..866S}
{Smith}, K.~L., {Shields}, G.~A., {Bonning}, E.~W., {et~al.} 2010, \apj, 716,
  866, \dodoi{10.1088/0004-637X/716/1/866}

\bibitem[{{Smith} {et~al.}(2012){Smith}, {Shields}, {Salviander}, {Stevens}, \&
  {Rosario}}]{2012ApJ...752...63S}
{Smith}, K.~L., {Shields}, G.~A., {Salviander}, S., {Stevens}, A.~C., \&
  {Rosario}, D.~J. 2012, \apj, 752, 63, \dodoi{10.1088/0004-637X/752/1/63}

\bibitem[{{Van Wassenhove} {et~al.}(2012){Van Wassenhove}, {Volonteri},
  {Mayer}, {Dotti}, {Bellovary}, \& {Callegari}}]{2012ApJ...748L...7V}
{Van Wassenhove}, S., {Volonteri}, M., {Mayer}, L., {et~al.} 2012, \apjl, 748,
  L7, \dodoi{10.1088/2041-8205/748/1/L7}

\bibitem[{{Villar Mart{\'{\i}}n} {et~al.}(2014){Villar Mart{\'{\i}}n},
  {Emonts}, {Humphrey}, {Cabrera Lavers}, \& {Binette}}]{2014MNRAS.440.3202V}
{Villar Mart{\'{\i}}n}, M., {Emonts}, B., {Humphrey}, A., {Cabrera Lavers}, A.,
  \& {Binette}, L. 2014, \mnras, 440, 3202, \dodoi{10.1093/mnras/stu448}

\bibitem[{{Wagner} {et~al.}(2012){Wagner}, {Bicknell}, \&
  {Umemura}}]{2012ApJ...757..136W}
{Wagner}, A.~Y., {Bicknell}, G.~V., \& {Umemura}, M. 2012, \apj, 757, 136,
  \dodoi{10.1088/0004-637X/757/2/136}

\bibitem[{{Wang} {et~al.}(2009){Wang}, {Chen}, {Hu}, {Mao}, {Zhang}, \&
  {Bian}}]{2009ApJ...705L..76W}
{Wang}, J.-M., {Chen}, Y.-M., {Hu}, C., {et~al.} 2009, \apjl, 705, L76,
  \dodoi{10.1088/0004-637X/705/1/L76}

\bibitem[{{Woo} {et~al.}(2014){Woo}, {Cho}, {Husemann}, {Komossa}, {Park}, \&
  {Bennert}}]{2014MNRAS.437...32W}
{Woo}, J.-H., {Cho}, H., {Husemann}, B., {et~al.} 2014, \mnras, 437, 32,
  \dodoi{10.1093/mnras/stt1846}

\bibitem[{{Xu} \& {Komossa}(2009)}]{2009ApJ...705L..20X}
{Xu}, D., \& {Komossa}, S. 2009, \apjl, 705, L20,
  \dodoi{10.1088/0004-637X/705/1/L20}

\bibitem[{{Zhou} {et~al.}(2004){Zhou}, {Wang}, {Zhang}, {Dong}, \&
  {Li}}]{2004ApJ...604L..33Z}
{Zhou}, H., {Wang}, T., {Zhang}, X., {Dong}, X., \& {Li}, C. 2004, \apjl, 604,
  L33, \dodoi{10.1086/383310}

\end{thebibliography}
%%%%%%%%%%%%%%%%%%%%%%%%%%%%%%%%%%%%%%%%%%%%%%%%%%%%%%%%%

%\label{lastpage}

\end{document}